\documentclass[11pt,a4paper]{article}
\usepackage[margin=25mm]{geometry}
\usepackage[T1]{fontenc}
\usepackage{lmodern}
\usepackage{type1cm}
\usepackage{graphicx}
\usepackage{siunitx}
\usepackage{amsmath}
\usepackage{amssymb}
\usepackage{booktabs}
\usepackage{xcolor}
\usepackage{placeins}   %
\usepackage[title,titletoc]{appendix}
\usepackage[numbers,sort&compress]{natbib}
\usepackage[hidelinks]{hyperref}
\usepackage{microtype}
\usepackage[font=footnotesize,labelfont=bf]{caption}
\graphicspath{{figures/}}
\setkeys{Gin}{width=0.88\linewidth,keepaspectratio}  %

\newcommand{\claim}[2]{#2}                       %

\newcommand{\EqSolver}{TokaMaker}
\newcommand{\EqKind}{fixed-boundary}
\newcommand{\EqGrid}{$129\times129$}

\newcommand{\qff}{q_{95}}
\newcommand{\Wth}{W_{\mathrm{th}}}
\newcommand{\Ip}{I_{\mathrm p}}
\newcommand{\Pnbi}{P_{\mathrm{NBI}}}

\title{Model-resilient multivariable tokamak feedback control using reinforcement learning}
\makeatletter
\hypersetup{pdftitle={\@title}}
\makeatother
\author{JangWoo Choi\textsuperscript{1}, Hyungkeun Nam\textsuperscript{1}, Jaemin Seo\textsuperscript{1*}\\[4pt]
\small \textsuperscript{1}Department of Physics, Chung-Ang University, Seoul, Republic of Korea\\
\small \textsuperscript{*}Corresponding author: \href{mailto:jseo@cau.ac.kr}{jseo@cau.ac.kr}}
\date{}

\begin{document}
\maketitle

\begin{abstract}
Sustaining high-performance plasmas in nuclear fusion reactors requires simultaneous feedback control of multiple plasma parameters to their desired values. However, adjusting individual actuators in tokamaks induces coupled responses in multiple plasma parameters, necessitating coordinated feedback control. To enable intelligent coordination of these shared actuators, we employ reinforcement learning. In this work, we develop a single reinforcement-learning policy that jointly controls the stored energy and safety factor of a plasma in an integrated tokamak simulation environment. The policy coordinates plasma current, heating power, and elongation in response to plasma observations and target values. In particular, the trained policy still responds appropriately to artificially introduced measurement biases and maintains robust control under changes in the underlying physics model. These results indicate that the policy has learned to flexibly regulate the plasma state through observation-based feedback at each control step, rather than simply memorising control trajectories associated with given targets. This suggests that a controller trained entirely in simulation could potentially be transferred to real tokamaks, where measurement noise, biases, and uncertainties in the underlying physics are present. This simulation-based approach offers a route towards developing controllers for future devices such as ITER and for devices with limited experimental data.

\end{abstract}

\section{Introduction}
\label{sec:intro}

Tokamaks use magnetic fields to confine a hot plasma in a toroidal configuration. Initiating a
discharge and maintaining the desired operating state require plasma current, shape and thermal
conditions to be controlled according to the discharge phase. The plasma control system coordinates
multiple actuators so that different physical quantities track their respective
targets~\cite{humphreys2015,snipes2014physics}. Because an actuator can affect several quantities,
controller design must specify which actuators are adjusted to bring each quantity towards its
target~\cite{walker2020introduction}. For example, a proportional--integral--derivative (PID) shape
controller on DIII-D adjusted shaping-coil voltages in response to flux errors at boundary control
points and to X-point position errors~\cite{humphreys2000shape}. The edge safety factor $\qff$ is
used to assess discharge operating limits associated with magnetohydrodynamic stability~\cite{piovesan2014lowq}. Thermal
stored energy $\Wth$ characterises the plasma's overall thermal state and serves as a performance
indicator~\cite{iterphysicsbasis1999}.

Multiple-input, multiple-output (MIMO) control is difficult because several controlled quantities
depend on the same actuators and the plasma response to each actuator varies with the operating
point~\cite{walker2020introduction,kim2012robust}. On DIII-D, joint control of stored energy and
toroidal rotation required coordinated changes in neutral-beam power~\cite{scoville2007rotation}.
In feedback loops for individual quantities, control gains are tuned over repeated discharges or
adjusted within a discharge as the plasma response
changes~\cite{eldon2017detachment,lennholm1997adaptive}. However, changing the gain in one loop can
alter the response of other controlled quantities through shared actuators, making independent
tuning difficult. As the numbers of controlled quantities and actuators increase, more gains and
coupling relationships must be considered together. On JET, a shape controller used a singular value
decomposition of the linearised response matrix to identify combinations of plasma--wall gaps
for feedback~\cite{ambrosino2008outputregulation,ariola2005jetshape}. On DIII-D, a controller based
on an experimentally identified response model regulated the poloidal flux profile and normalised
beta together~\cite{moreau2013integrated}. When loops request different actions
from the same actuator, the final command is chosen according to control priorities and the
available actuator range~\cite{maljaars2017allocation}. Joint tracking therefore requires a feedback
law that coordinates actuator actions according to the measured plasma state and multiple targets.

Reinforcement learning (RL) provides a way to learn a feedback law through interaction with a
controlled system~\cite{sutton1992adaptive}. A policy selects actions from observations and learns
from the resulting state changes and rewards~\cite{kaelbling1996rl}. In a tokamak simulator, the
plasma model supplies the response to each combination of actuator commands, allowing a single
policy to learn how to coordinate the actuators.
Tokamak applications of RL include magnetic control of plasma shape on TCV~\cite{degrave2022},
control of tearing-instability risk on DIII-D~\cite{seo2024}, and performance control through
joint regulation of the safety-factor profile and normalised beta in JT-60SA
simulations~\cite{wakatsuki2023simultaneous}. These studies show
that RL policies can meet a range of control objectives in tokamak experiments and simulations.
However, during an actual experiment, a tracking error can arise from either a shift in a measurement or a change in the plasma response to the same
actuators. The transfer of a policy learned in simulations to experiments requires a more careful understanding how a persistent bias in one measurement affects
actuator commands at fixed targets and how a change in the plasma response affects the range of
targets a single policy can track.

This study examines whether a policy adjusts its actuators in response to changes in the measured
plasma state and over what range of targets it can maintain joint tracking when the plant changes.
A single policy trained in an ITER-scale tokamak simulation was evaluated without retraining,
with uncertainty in plasma parameters and applied actuator values throughout the tests.
Tracking was assessed across fixed targets and from different initial states. The moving target
test then varied the targets during operation. In the measurement bias test, a bias was added to the stored energy measurement
supplied to the policy. This tested whether the policy
adjusted the actuators to bring that measurement back towards its target. The plant shift test
changed the plant configuration, including transport and heating conditions, and assessed joint
tracking inside and outside the training target domain. These tests assess the potential for a
simulation-trained policy to serve as a starting point for multivariable control on real tokamak experiments with
limited discharge data.

\FloatBarrier
\section{Simulation environment}\label{sec:env}

\label{sec:env:arch}\label{sec:env:model}
\begin{figure}[!t]
\centering
\includegraphics[width=\linewidth]{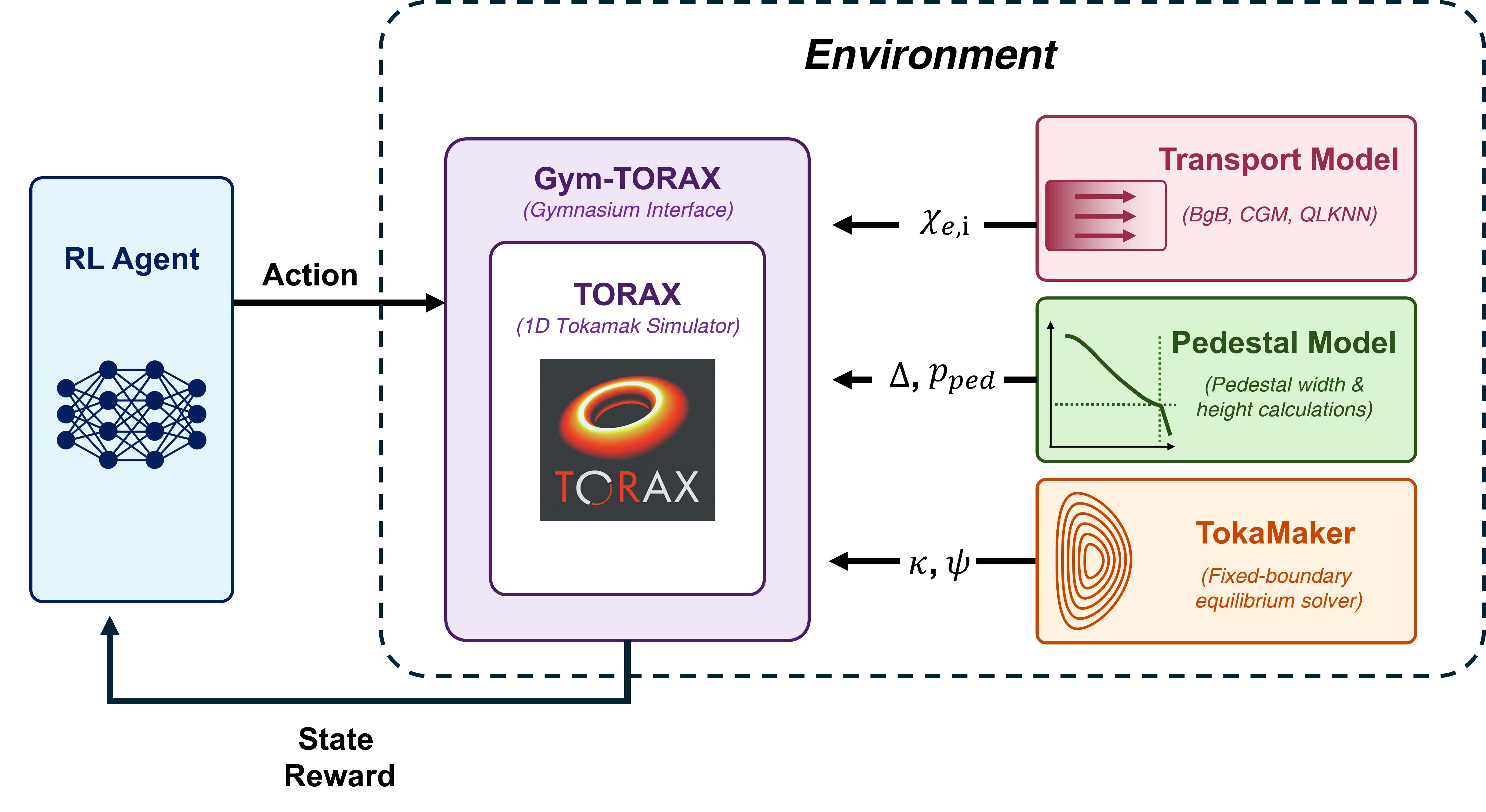}
\caption{Information flow between the simulation environment and the policy. Through Gym-TORAX, the policy adjusts the three actuators: plasma current, beam power and elongation. TORAX evolves the plasma state using heat diffusivities from the transport model, pedestal boundary conditions and \EqKind{} magnetic geometry provided by \EqSolver{}. The computed thermal stored energy $\Wth$ and safety factor $\qff$ are included in the policy's next observation. Training uses BgB, and the frozen policy is also evaluated with CGM and QLKNN.}
\label{fig:plant_loop}
\end{figure}

Policy training and evaluation use a simulation environment for an ITER-scale deuterium--tritium
H-mode plasma. The simulation starts in H-mode and evolves the plasma in response to plasma current,
neutral beam power and elongation. TORAX computes core plasma transport~\cite{torax}, and a
scaling-based pedestal model provides the boundary
conditions for the core profile~\cite{onjun2002}. The \EqSolver{} \EqKind{} equilibrium solver supplies the magnetic
geometry required for the transport calculation~\cite{hansen2024tokamaker}, and Gym-TORAX connects
the simulation environment to the RL policy~\cite{mouchamps2026gymtorax,gymnasium}. At each control
cycle, the policy adjusts the three actuators and the environment calculates the resulting plasma
response. The computed thermal stored energy $\Wth$ and edge safety factor $\qff$ are included in
the observations for the next control cycle. Figure~\ref{fig:plant_loop} shows this information
flow.

\FloatBarrier

\label{sec:env:transport}
TORAX evolves the ion temperature $T_i$, electron temperature $T_e$ and poloidal flux $\psi$ in the
normalised toroidal flux coordinate $\hat\rho$~\cite{torax}. Ion and electron energy transport are
described by their respective flux-surface-averaged heat transport equations, and the poloidal flux
evolves according to the current diffusion equation.
Equations~(\ref{eq:transport_i})--(\ref{eq:transport_psi}) give the ion heat transport, electron
heat transport and current diffusion equations, respectively.

\begin{equation}\label{eq:transport_i}
\begin{aligned}
\frac{3}{2}V'^{-5/3}
\left(\frac{\partial}{\partial t}-\frac{\dot\Phi_b}{2\Phi_b}
\frac{\partial}{\partial\hat\rho}\hat\rho\right)
\left[V'^{5/3}n_iT_i\right]
={}&\frac{1}{V'}\frac{\partial}{\partial\hat\rho}
\left[\chi_i n_i\frac{g_1}{V'}\frac{\partial T_i}{\partial\hat\rho}
-g_0q_i^{\mathrm{conv}}T_i\right]+Q_i.
\end{aligned}
\end{equation}

\begin{equation}\label{eq:transport_e}
\begin{aligned}
\frac{3}{2}V'^{-5/3}
\left(\frac{\partial}{\partial t}-\frac{\dot\Phi_b}{2\Phi_b}
\frac{\partial}{\partial\hat\rho}\hat\rho\right)
\left[V'^{5/3}n_eT_e\right]
={}&\frac{1}{V'}\frac{\partial}{\partial\hat\rho}
\left[\chi_e n_e\frac{g_1}{V'}\frac{\partial T_e}{\partial\hat\rho}
-g_0q_e^{\mathrm{conv}}T_e\right]+Q_e.
\end{aligned}
\end{equation}

\begin{equation}\label{eq:transport_psi}
\begin{aligned}
\frac{16\pi^2\sigma_\parallel\mu_0\hat\rho\,\Phi_b^2}{F^2}
\left(\frac{\partial\psi}{\partial t}-\frac{\hat\rho\dot\Phi_b}{2\Phi_b}
\frac{\partial\psi}{\partial\hat\rho}\right)
={}&\frac{\partial}{\partial\hat\rho}
\left(\frac{g_2g_3}{\hat\rho}\frac{\partial\psi}{\partial\hat\rho}\right)
-\frac{8\pi^2V'\mu_0\Phi_b}{F^2}
\langle\mathbf{B}\!\cdot\!\mathbf{j}_{\mathrm{ni}}\rangle .
\end{aligned}
\end{equation}

Here, $V'=\mathrm{d}V/\mathrm{d}\hat\rho$ is the derivative of volume with respect to the flux
coordinate, and $g_0$--$g_3$ are flux-surface geometric factors. The quantities $n_{i,e}$,
$\chi_{i,e}$, $q_{i,e}^{\mathrm{conv}}$ and $Q_{i,e}$ denote the ion and electron densities, heat
diffusivities, heat-convection terms and volumetric heat sources, respectively. The toroidal flux
enclosed by the plasma boundary is $\Phi_b$, $F=RB_\phi$ is a flux function associated with the
toroidal magnetic field, $\sigma_\parallel$ is the electrical conductivity parallel to the magnetic
field, and $\langle\mathbf{B}\cdot\mathbf{j}_{\mathrm{ni}}\rangle$ represents the non-inductive
current contribution. The heat sources and losses include neutral beam heating, fusion heating,
ohmic heating, ion--electron energy exchange, bremsstrahlung and cyclotron radiation. The heat flux
terms are calculated using the Bohm/gyro-Bohm (BgB) transport model~\cite{bgb1998} in the training
plant, and the critical-gradient model (CGM) and QLKNN for evaluations with modified plants. The ion
temperature-gradient threshold in CGM follows the model of Guo and Romanelli~\cite{cgm1993}, and
QLKNN is a neural network model that approximates transport predictions from the quasilinear
gyrokinetic code QuaLiKiz~\cite{vandeplassche2020qlknn,citrin2017qualikiz}.

\label{sec:env:pedestal}
The pedestal model specifies the pressure and position of the pedestal top for the core transport
calculation. The model adapts the pedestal-width scaling~\cite{onjun2002}, using the global poloidal beta in
$\Delta\propto\beta_{\mathrm p}^{\claim{L254}{0.5}}$. The pressure follows
$p_{\mathrm{ped}}\propto\Delta\,\alpha_c(\kappa)\,\kappa^m/q_{95}^{2}$, where $\alpha_c$ is the
ballooning-limit pressure gradient at the prescribed magnetic shear and triangularity, and the
elongation exponent $m$ and proportionality constant are specified for each plant configuration. The
electron density follows a prescribed density profile with a change in gradient at the pedestal top.
At each computational step, this profile is normalised so that the line-averaged density is a
prescribed fraction of the Greenwald density~\cite{greenwald1988limit,greenwald2002density}. Since the Greenwald
density is proportional to the plasma current, changes in current also affect the thermal state
through the density. Particle transport and fuelling are not evolved separately.

\label{sec:env:equilibrium}\label{sec:env:actuators}\label{sec:env:actmodel}
The three actuators in this simulation are the plasma current $\Ip$, neutral beam power $\Pnbi$ and
elongation $\kappa$; their permitted ranges and maximum changes per control cycle are listed in
Table~\ref{tab:interface}. The current affects flux evolution, equilibrium and the density profile,
while elongation affects the plasma boundary and pedestal model; beam power scales a Gaussian
heating profile with a fixed spatial shape. Physical actuator dynamics and lower-level current and
shape controllers are not modelled separately. \EqSolver{} receives the current and elongation,
together with the on-axis pressure from the preceding TORAX state. The resulting equilibrium
supplies the magnetic geometry for TORAX. The magnetic geometry is updated every
\claim{L202}{\SI{5}{\second}} and held fixed between updates, while TORAX continues to evolve the
poloidal flux. The $\qff$ used for control is calculated from the flux evolved by TORAX on this
geometry. For the equilibrium calculation, the plasma boundary is prescribed in the Miller form
given in equation~(\ref{eq:boundary})~\cite{miller1998}, where $\delta$ controls the boundary
triangularity and $\kappa$ sets the elongation.

\begin{equation}\label{eq:boundary}
R_b(\theta)=R_0+a\cos(\theta+\delta\sin\theta),\qquad
Z_b(\theta)=Z_0+\kappa a\sin\theta ,
\end{equation}

\begin{figure}[!t]
\centering
\includegraphics[width=0.80\linewidth]{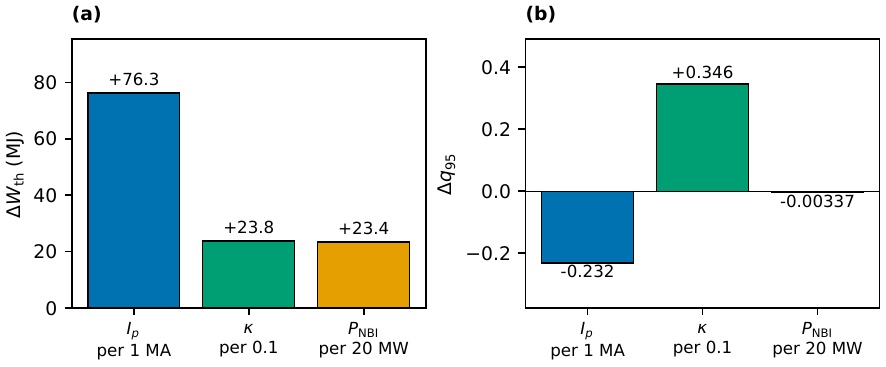}
\caption{Open-loop sensitivities of the settled (a) thermal stored energy and (b) edge safety factor to the three actuators in the training plant. The reference settings are \claim{L264}{$\Ip=\SI{15}{\mega\ampere}$, $\kappa=1.88$ and $\Pnbi=\SI{50}{\mega\watt}$}. Each scan uses nominal plasma parameters and applies the commanded actuator values directly, varying one actuator while holding the other two fixed. Each bar shows the average output change across the scan range for a change of \claim{L265}{\SI{1}{\mega\ampere}} in current, \claim{L266}{0.1} in elongation or \claim{L267}{\SI{20}{\mega\watt}} in beam power.}
\label{fig:coupling}
\end{figure}

\label{sec:env:coupling}
Open-loop scans of the training plant are used to characterise actuator--output sensitivities and
coupling. The scans use nominal plasma parameters and apply the actuator commands directly.
With the policy disabled, one actuator is varied while the other two are held at their reference values, and the changes in thermal stored energy and safety factor
are measured after settling. Figure~\ref{fig:coupling} shows the average output changes for the
specified actuator increments across each scan range. Within these ranges, increasing the current
raises the thermal stored energy and lowers the safety factor, whereas increasing the elongation
raises both quantities. Changes in beam power primarily affect the thermal stored energy and have a
comparatively small effect on the safety factor. To account for this coupling, joint tracking of
$\Wth$ and $\qff$ is formulated as a multiple-input, multiple-output control problem in which all
three actuators are adjusted together.

\FloatBarrier

\label{sec:env:variations}
During training and closed-loop evaluation, the simulation introduces uncertainty in plasma
parameters such as the effective charge and in the values applied to all three actuators. Bounded random walks updated every
\claim{L272}{\SI{2}{\second}} vary the plasma parameters about their nominal values and the applied actuator values
about the values requested by the policy. For example, the applied neutral-beam power can fluctuate
even when the power command is held fixed, changing the heating delivered to the plasma. Under
these conditions, the policy is trained to adjust the actuators according to changing plasma observations
rather than simply memorise an actuator combination for each target. The imposed deviations are not
supplied directly to the policy, and no additional noise is added to the measurements of the plasma outputs.

The CGM and QLKNN evaluations include uncertainty and use nominal plant configurations that
differ from the training configuration. The transport model, pedestal parameters, prescribed density
profile and beam absorption fraction are changed together. Changing the beam absorption fraction
alters the heating delivered to the plasma at the same beam power; changes in the pedestal and
density settings also alter the thermal response to current, elongation and heating. The same policy
is applied without retraining to assess joint tracking in the modified plants. The device and plasma
conditions, detailed model settings, numerical methods and uncertainty parameters are summarised
in Appendix~\ref{sec:appendix:plant}.

\FloatBarrier
\section{Reinforcement-learning control method}\label{sec:method}

\subsection{Controller formulation and training}
\label{sec:method:formulation}\label{sec:method:reward}\label{sec:method:training}

The controller regulates the thermal stored energy $\Wth$ and edge safety factor $\qff$ by changing
the plasma current $\Ip$, neutral-beam power $\Pnbi$ and elongation $\kappa$. Reinforcement learning
formulates this task as repeated interaction between a policy and the simulation
environment~\cite{kaelbling1996rl}. At control step $k$, the policy $\pi_\theta(a_k\mid o_k)$ selects
an action $a_k$ from the observation $o_k$ and receives a reward $r_k$ after the plant advances.
Training adjusts the neural-network parameters $\theta$ to maximise the expected discounted return
$\mathbb{E}[\sum_k\gamma^k r_k]$, where $\gamma$ weights future rewards. The resulting policy is
used as a fixed feedback law during evaluation.

The observation contains \claim{L210}{seven} scalars,
\begin{equation}\label{eq:obs}
o_k=\bigl(\Ip,\ \Pnbi,\ \kappa,\ \Wth^{\star},\ \Wth,\ \qff^{\star},\ \qff\bigr),
\end{equation}
comprising the actuator setpoints, measured outputs and target pair $(\Wth^\star,\qff^\star)$.
The policy does not directly observe the plasma profiles, the imposed deviations or time. Its action
specifies one increment per actuator. Each setpoint is the running sum of these increments, restricted
to the actuator range and maximum change per control period in table~\ref{tab:interface}. The policy
therefore receives the current values of the setpoints it has adjusted, as well as the plasma response.

\begin{table}[!htbp]
\caption{Observation and action spaces of the policy. The observation of equation~(\ref{eq:obs}) consists of the three actuator setpoints, the two targets and the two measured outputs; the action is one increment per actuator, added to the setpoint at each control period.}
\label{tab:interface}
\centering
\small
\renewcommand{\arraystretch}{1.35}
\begin{tabular}{l l c c}
\toprule
Quantity & Description & Min & Max \\
\midrule
\multicolumn{4}{l}{\textit{Observation space}} \\
\(\Ip\) & plasma current & \claim{L98}{10 MA} & \claim{L98}{16 MA} \\
\(\Pnbi\) & neutral-beam power & \claim{L98}{0 MW} & \claim{L98}{100 MW} \\
\(\kappa\) & elongation & \claim{L98}{1.65} & \claim{L98}{2.00} \\
\(\Wth^\star\) & stored-energy target & \claim{L250}{150 MJ} & \claim{L250}{448.5 MJ} \\
\(\Wth\) & measured stored energy & -- & -- \\
\(\qff^\star\) & safety-factor target & \claim{L250}{2.424} & \claim{L250}{5.604} \\
\(\qff\) & measured safety factor & -- & -- \\
\midrule
\multicolumn{4}{l}{\textit{Action space}} \\
\(\Delta\Ip\) & plasma-current increment & \claim{L98}{$-$0.05 MA} & \claim{L98}{+0.05 MA} \\
\(\Delta\Pnbi\) & neutral-beam-power increment & \claim{L98}{$-$5 MW} & \claim{L98}{+5 MW} \\
\(\Delta\kappa\) & elongation increment & \claim{L98}{$-$0.01} & \claim{L98}{+0.01} \\
\bottomrule
\end{tabular}
\end{table}

\begin{figure}[!t]
\centering
\includegraphics[width=0.80\linewidth]{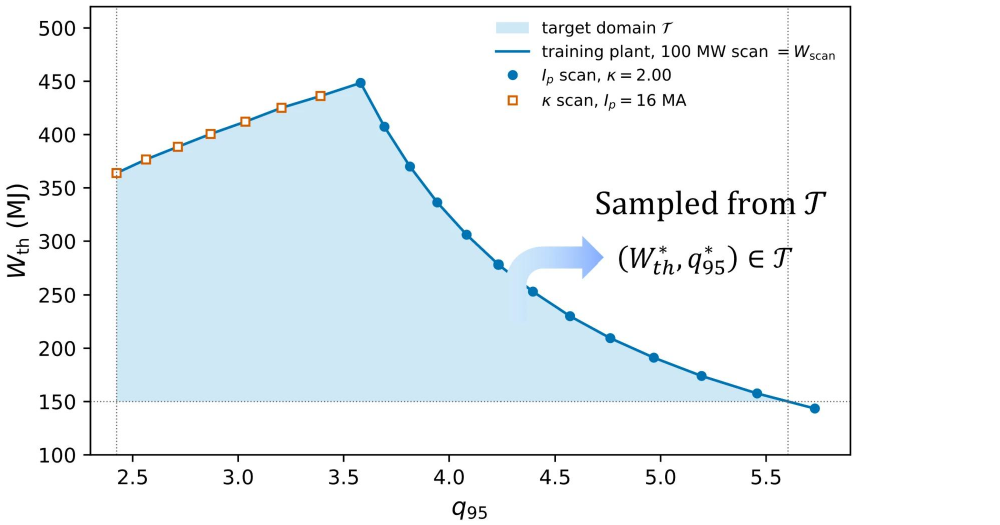}
\caption{Settled stored energy and edge safety factor from the open-loop current and elongation scans of the training plant at \claim{L200}{\SI{100}{\mega\watt}}. Solid line, interpolated sampling boundary; shaded area, target domain $\mathcal{T}$ of equation~(\ref{eq:domain}). The arrow marks the sampling of a target pair $(\Wth^\star,\qff^\star)$ from $\mathcal{T}$ for a training episode.}
\label{fig:plant_map_transfer}
\end{figure}

\label{sec:env:targets}\label{sec:method:targets}
The training target domain is defined by an open-loop scan of the training plant at
\claim{L200}{\SI{100}{\mega\watt}} of beam power. Writing the interpolated upper boundary of the
settled scan states as $W_{\rm scan}(\qff)$ gives
\begin{equation}\label{eq:domain}
\mathcal{T}=\left\{(\Wth^{\star},\qff^{\star}):
\claim{L250}{2.424\le\qff^{\star}\le5.604},\quad
\claim{L250}{\SI{150}{\mega\joule}\le\Wth^{\star}\le W_{\rm scan}(\qff^{\star})}\right\}.
\end{equation}
Targets are sampled conditionally: $\qff^\star$ is drawn first, and the corresponding scan boundary
sets the upper limit for $\Wth^\star$. Figure~\ref{fig:plant_map_transfer} shows the resulting
training-request region.

\FloatBarrier

Each training episode lasts \claim{L96}{\SI{200}{\second}}, comprising \claim{L96}{400} control
steps of \claim{L96}{\SI{0.5}{\second}}. Episodes use fixed, stepped and random-walk targets with
probabilities \claim{L92}{0.4, 0.3 and 0.3}, respectively. A failed transport solve or stored
energy outside \claim{L211}{\SIrange{100}{700}{\mega\joule}} terminates the episode.

The reward combines the errors in both controlled outputs,
\begin{equation}\label{eq:reward}
r=\prod_{c\,\in\,\{W,q\}}\frac{2}{1+\exp\!\left(k_c\lvert e_c\rvert\right)},
\qquad e_W=\Wth^\star-\Wth,\quad e_q=\qff^\star-\qff .
\end{equation}
Each factor decreases monotonically with the absolute tracking error. It equals one at zero error
and \claim{L90}{0.1} at \claim{L90}{\SI{200}{\mega\joule}} of stored-energy error or
\claim{L90}{2.0} of safety-factor error; these values set the slopes $k_W$ and $k_q$. Their
product rewards simultaneous reduction of both errors. A failed step receives the reward
\claim{L212}{$-1$}.

The policy was trained with proximal policy optimisation (PPO)~\cite{ppo} using
Stable-Baselines3~\cite{sb3}. PPO collects trajectories with the current policy and updates an actor
network, which selects actions, and a critic network, which estimates future return. Its update
rule discourages large policy changes between successive training cycles. The learning algorithm uses
the collected observations, actions and rewards without constructing an internal model of the
plant dynamics. Observations are standardised using running training statistics. The policy was trained under uncertainty
with a budget of \claim{L321}{4000} simulated episodes. Table~\ref{tab:training} gives the network
and optimisation settings. The reported policy comes
from a single training run.

\subsection{Evaluation protocol}\label{sec:method:evaluation}

\claim{L290}{Each test case consists of one episode under uncertainty. Evaluation uses deterministic actions
with the policy parameters and observation-standardisation statistics frozen.}
Deterministic actions use the mean of the policy's action distribution. A terminal value is the
mean over the last \claim{L101}{20} control steps, or \claim{L101}{\SI{10}{\second}}. The signed
relative tracking errors are
\begin{equation}\label{eq:eps}
\epsilon_W=\frac{\Wth^\star-\Wth}{\Wth^\star},\qquad
\epsilon_q=\frac{\qff^\star-\qff}{\qff^\star},
\end{equation}
with positive values indicating outputs below their targets. Terminal errors use the terminal
mean of each output. A target is counted as met when both absolute relative errors are within
\claim{L104}{5}\,\%; this tolerance is an analysis convention. Where a terminal mean does not
describe whether the target is held, the fraction of the terminal window within tolerance and the
largest excursion are also reported. \claim{L312}{All evaluation episodes reached the horizon.}

In the measurement-bias test, an offset is added only to the stored energy supplied to the policy.
The target and plant equations are unchanged, and the true outputs are recorded independently of
the biased observation. The plant-change evaluations use a common target list and additional grids
extending outside $\mathcal{T}$.

\FloatBarrier
\section{Results}
\label{sec:results}

A feedback controller must adjust its actuators in response to changes in either the target or the measured plasma state. To evaluate the control performance under various conditions, we conducted four tests. The fixed-target tests assessed tracking accuracy across target pairs and compared trajectories to a common target from different initial states. The target-change test examined how the policy adjusted its actuators when $\Wth^\star$ changed while $\qff^\star$ remained fixed. The measurement-bias test held both targets fixed and examined how the actuator commands changed when a bias was added only to the $\Wth$ measurement supplied to the policy. With future tokamak applications in mind, the final simulations assessed how well the policy retained joint tracking without retraining when the transport model and plasma conditions differed from those used in training.

All tests used the same policy without retraining and included uncertainty in plasma parameters and applied actuator values. Terminal errors were calculated from the outputs averaged over the final \claim{L101}{\SI{10}{\second}}, whereas final actuator setpoints were taken from the last control step. A case met the joint tracking criterion if both outputs had absolute relative terminal errors of \claim{L104}{5}\,\% or less.

\subsection{Fixed target test}
\label{sec:results:nominal}

A single trained policy was evaluated on \claim{L300}{100} different fixed target pairs sampled from the training target domain $\mathcal{T}$. Table~\ref{tab:distribution} shows the tracking error metrics for different-target controls with a single trained RL policy. Across both outputs and all target pairs, the relative terminal errors gave a root-mean-square error (RMSE) of \claim{L350}{3.17}\,\%. \claim{L301}{The policy thus tracked both outputs accurately for most of the sampled target pairs}.

\begin{table}
\caption{Terminal tracking errors on the training plant for \claim{L300}{100} target pairs sampled from the
target domain, one episode each. $|\Delta\Wth|=|\Wth^\star-\Wth|$ and $|\Delta\qff|=|\qff^\star-\qff|$ are the
absolute terminal errors, $\epsilon_W$ and $\epsilon_q$ the relative errors (equation~(\ref{eq:eps})), and
$\langle\cdot\rangle$ the mean over the targets; RMSE $=\sqrt{\langle(\epsilon_W^2+\epsilon_q^2)/2\rangle}$.}
\centering
\begin{tabular}{l c}
\toprule
Quantity & Value \\
\midrule
$\langle|\epsilon_W|\rangle$ (\%) & \claim{L301}{1.6} \\
$\langle|\epsilon_q|\rangle$ (\%) & \claim{L301}{1.9} \\
$\langle|\Delta\Wth|\rangle$ (MJ) & \claim{L345}{4.1} \\
$\langle|\Delta\qff|\rangle$ & \claim{L345}{0.065} \\
RMSE (\%) & \claim{L350}{3.17} \\
Targets with $|\epsilon_W|$ and $|\epsilon_q|$ within \claim{L104}{5}\,\% & \claim{L301}{85/100} \\
\bottomrule
\end{tabular}
\label{tab:distribution}
\end{table}

\claim{L303}{Joint tracking failures were more common at higher stored-energy targets}, and \claim{L356}{usually only the safety factor was outside tolerance}. \claim{L356}{In most failed runs, both current and beam power ended at their upper limits}. \claim{L265}{The current could not be raised further to lower an above-target safety factor}. \claim{L266}{Reducing elongation could lower both outputs (figure~\ref{fig:coupling}), but no additional heating was available to offset the drop in stored energy}. With the other output on target at these operating points, \claim{L368}{an error of a given relative magnitude reduces the reward more for $\Wth$ than for $\qff$ because the two reward factors are scaled differently}. The policy thus appears to have sacrificed some safety-factor accuracy to keep stored energy near target and earn the higher reward.

\subsubsection{Tracking from different initial states}
\label{sec:results:initial}

\begin{figure}[!t]
\centering
\includegraphics{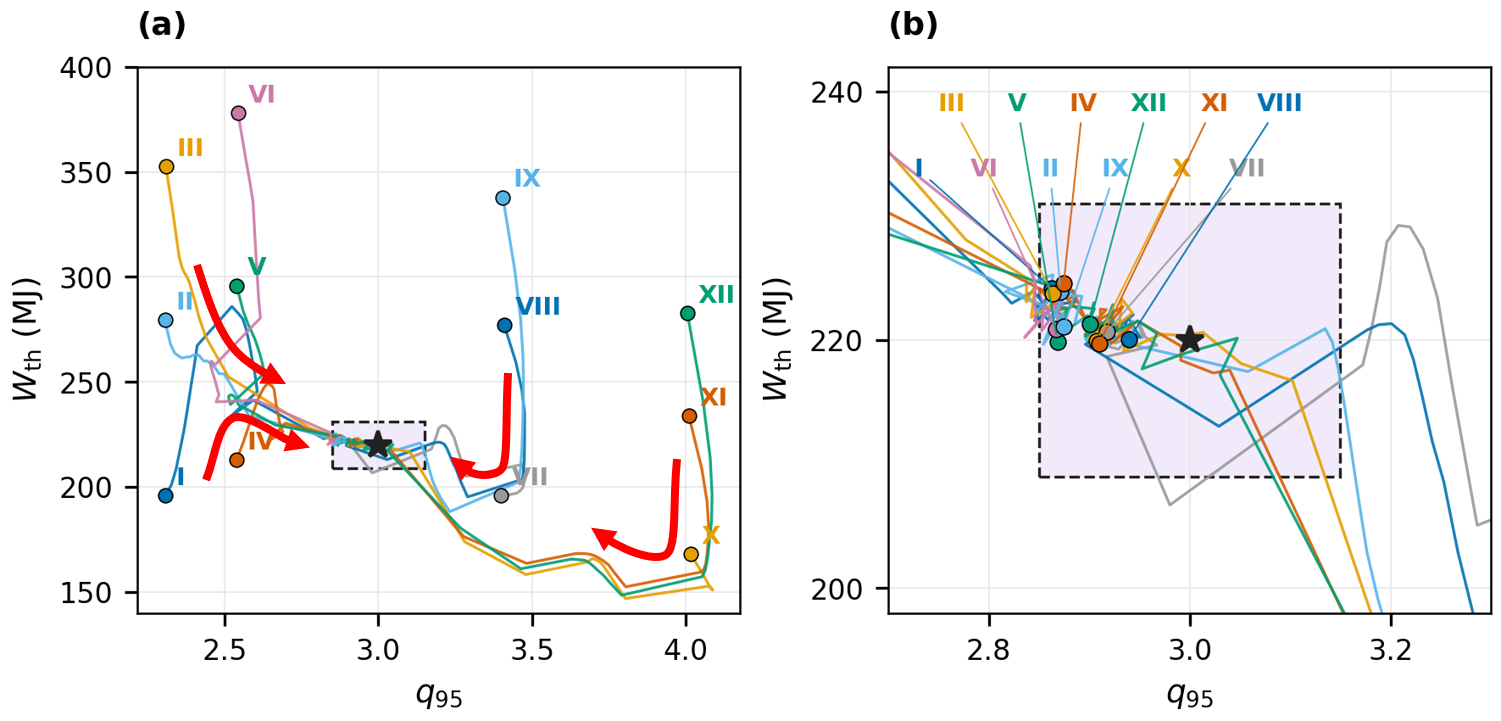}
\caption{Trajectories from \claim{L352}{twelve} initial states towards the common target \claim{L352}{$(\Wth^\star,\qff^\star)=(\SI{220}{\mega\joule},3.0)$} on the training plant. (a) Full trajectories. Red arrows indicate the direction of approach. Circles labelled I--XII mark the initial states and the star marks the target. (b) Trajectories near the target. Circles mark the final states. Inside the shaded box, each output is within \claim{L104}{5}\,\% of its own target.}
\label{fig:nominal_tracking}
\end{figure}

In tokamak operation, the plasma state at the start of control can vary between discharges even when the target operating point is the same. The target pair for this test was \claim{L352}{$(\Wth^\star,\qff^\star)=(\SI{220}{\mega\joule},3.0)$}. Figure~\ref{fig:nominal_tracking} shows how the same policy approached this target from \claim{L352}{12} initial states with different combinations of $\Wth$ and $\qff$. For example, in case III, reaching the target required a decrease in $\Wth$ and an increase in $\qff$, whereas case VII required the opposite changes. All \claim{L353}{12} cases met the joint tracking criterion.

\begin{table}[!t]
\caption{Tracking the common target \claim{L352}{$(\Wth^\star,\qff^\star)=(\SI{220}{\mega\joule},3.0)$} from the \claim{L352}{twelve} initial states in figure~\ref{fig:nominal_tracking}. The table gives the initial stored energy, plasma current and elongation, the signed relative terminal errors in the two outputs (\%), and the final actuator values. Terminal errors use the mean outputs over the final \claim{L101}{\SI{10}{\second}}. Actuator values are taken at the final control step.}
\centering
\small
\begin{tabular}{l c c c c c c c c}
\toprule
Label & $W_0$ (MJ) & $I_{p,0}$ (MA) & $\kappa_0$ & $\epsilon_W$ & $\epsilon_q$ & $\Ip$ (MA) & $\Pnbi$ (MW) & $\kappa$ \\
\midrule
I & \claim{L354}{196} & \claim{L354}{15.9} & \claim{L354}{1.66} & \claim{L354}{-1.62} & \claim{L354}{+4.94} & \claim{L354}{15.08} & \claim{L354}{13.2} & \claim{L354}{1.728} \\
II & \claim{L354}{279} & \claim{L354}{16.0} & \claim{L354}{1.65} & \claim{L354}{-1.24} & \claim{L354}{+3.94} & \claim{L354}{15.21} & \claim{L354}{22.3} & \claim{L354}{1.726} \\
III & \claim{L354}{352} & \claim{L354}{16.0} & \claim{L354}{1.65} & \claim{L354}{-1.34} & \claim{L354}{+4.57} & \claim{L354}{15.22} & \claim{L354}{16.7} & \claim{L354}{1.732} \\
IV & \claim{L354}{213} & \claim{L354}{15.9} & \claim{L354}{1.76} & \claim{L354}{-1.42} & \claim{L354}{+4.27} & \claim{L354}{15.35} & \claim{L354}{18.2} & \claim{L354}{1.716} \\
V & \claim{L354}{296} & \claim{L354}{16.0} & \claim{L354}{1.74} & \claim{L354}{-0.59} & \claim{L354}{+4.08} & \claim{L354}{14.54} & \claim{L354}{34.0} & \claim{L354}{1.752} \\
VI & \claim{L354}{378} & \claim{L354}{16.0} & \claim{L354}{1.74} & \claim{L354}{-0.77} & \claim{L354}{+4.53} & \claim{L354}{14.49} & \claim{L354}{30.4} & \claim{L354}{1.755} \\
VII & \claim{L354}{196} & \claim{L354}{13.4} & \claim{L354}{1.87} & \claim{L354}{+0.30} & \claim{L354}{+2.60} & \claim{L354}{14.72} & \claim{L354}{44.4} & \claim{L354}{1.765} \\
VIII & \claim{L354}{277} & \claim{L354}{13.4} & \claim{L354}{1.86} & \claim{L354}{-0.28} & \claim{L354}{+2.20} & \claim{L354}{15.05} & \claim{L354}{40.3} & \claim{L354}{1.734} \\
IX & \claim{L354}{338} & \claim{L354}{13.4} & \claim{L354}{1.86} & \claim{L354}{-0.57} & \claim{L354}{+4.25} & \claim{L354}{14.63} & \claim{L354}{31.4} & \claim{L354}{1.758} \\
X & \claim{L354}{168} & \claim{L354}{11.4} & \claim{L354}{1.86} & \claim{L354}{-0.75} & \claim{L354}{+3.31} & \claim{L354}{15.22} & \claim{L354}{31.3} & \claim{L354}{1.728} \\
XI & \claim{L354}{234} & \claim{L354}{11.4} & \claim{L354}{1.86} & \claim{L354}{-0.35} & \claim{L354}{+3.01} & \claim{L354}{14.89} & \claim{L354}{37.0} & \claim{L354}{1.733} \\
XII & \claim{L354}{283} & \claim{L354}{11.4} & \claim{L354}{1.86} & \claim{L354}{-0.71} & \claim{L354}{+3.22} & \claim{L354}{14.97} & \claim{L354}{33.1} & \claim{L354}{1.728} \\
\bottomrule
\end{tabular}
\label{tab:initial_conditions}
\end{table}

The trajectories in figure~\ref{fig:nominal_tracking}(a) followed two broad paths depending on whether $\qff$ was initially below or above its target. In cases I--VI, the trajectories started at different values of $\Wth$ and drew together to the left of the target point as $\qff$ increased. In cases VII--XII, the trajectories generally passed below $\Wth^\star$ before approaching the target point as $\Wth$ rose and $\qff$ fell.

\begin{samepage}
Cases III and VI followed the same broad path and reached similar terminal values of $\Wth$ and $\qff$, although \claim{L354}{case VI ended with lower plasma current and higher beam power and elongation than case III}. Although the test specified only the target point and prescribed neither intermediate states nor a sequence of actuator commands, trajectories from different starting points followed two broad paths. This pattern suggests that the policy may have learned to steer the plasma onto one path or the other according to its initial state. Along each path, the policy could reuse a control strategy across cases, reducing the need to learn a separate sequence of actuator commands for every combination of initial state and target.\par
\end{samepage}

\FloatBarrier

\subsection{Moving target test}
\label{sec:results:step}

\begin{figure}[p]
\centering
\includegraphics[width=\linewidth,height=0.89\textheight,keepaspectratio]{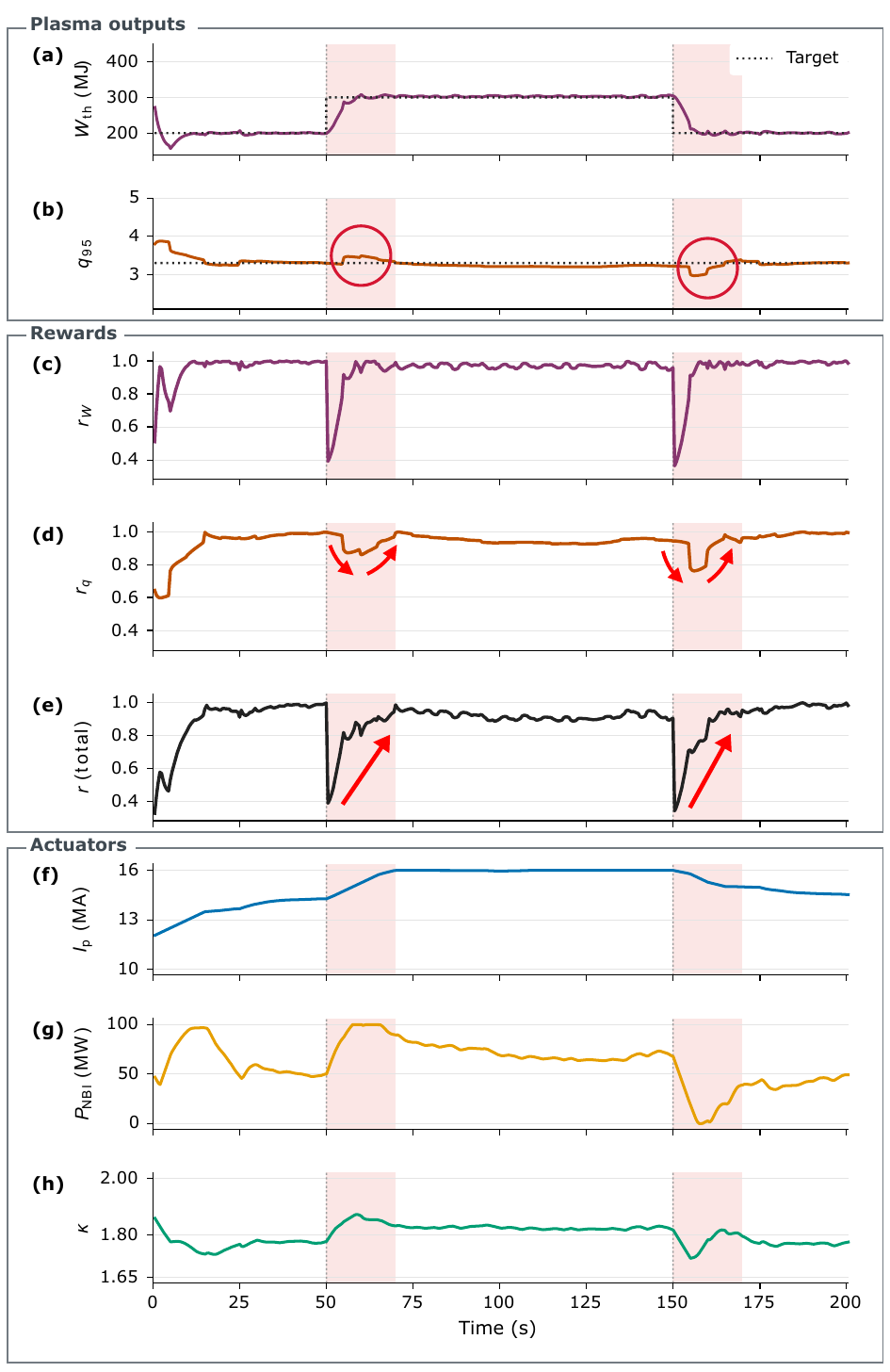}
\caption{Moving target test on the training plant. (a) Thermal stored energy $\Wth$ and its target (dotted), raised from \claim{L305}{\SI{200}{\mega\joule}} to \claim{L305}{\SI{300}{\mega\joule}} at \claim{L305}{\SI{50}{\second}} and returned to \claim{L305}{\SI{200}{\mega\joule}} at \claim{L305}{\SI{150}{\second}}. (b) Edge safety factor $\qff$ and its fixed target of \claim{L305}{3.3} (dotted). Circles mark the largest absolute $\qff$ error after each change. (c) Stored-energy reward factor $r_W$. (d) Safety-factor reward factor $r_q$, which falls briefly after each target change and then recovers. (e) Total reward $r=r_W r_q$ from equation~(\ref{eq:reward}). (f) Plasma current $\Ip$. (g) Neutral-beam power $\Pnbi$. (h) Elongation $\kappa$. Vertical dashed lines mark the target changes, and the shaded bands cover the first \claim{L358}{\SI{20}{\second}} after each change.}
\label{fig:step_body}
\end{figure}

In the moving-target test, the stored-energy target was modulated within an evaluation scenario, while the $\qff$ target remained fixed, as shown in figure~\ref{fig:step_body}. The policy was trained to maximise the expected discounted return, which gives greater weight to rewards received sooner. After each change, the policy drove stored energy rapidly towards its new target, \claim{L362}{with each of the three actuator commands changing at or near its maximum permitted rate during part of the response}. \claim{L360}{The safety-factor error peaked in magnitude when stored energy was close to target} and then decreased again. Interpreting this temporary deterioration and subsequent recovery in safety-factor tracking requires examining the total reward per step and both component factors.

Immediately after each target change, the stored-energy error increased, reducing both the stored-energy reward factor $r_W$ and the total reward per step, $r=r_W r_q$. \claim{L341}{The subsequent recovery in $r_W$ was fast enough for the product to rise even as the safety-factor reward factor $r_q$ continued to fall}. Once $\Wth$ was close to target, \claim{L361}{the policy adjusted plasma current and elongation in opposite directions} as $\qff$ moved back towards its fixed target. The sequence after both steps suggests that the policy learned to recover total reward early by restoring $r_W$ first at a temporary cost to $r_q$. This trade-off helps explain why, in \claim{L356}{most fixed-target cases that missed the joint criterion}, the policy kept stored energy within tolerance while the safety factor remained outside it (section~\ref{sec:results:nominal}).

\subsection{Measurement bias test}
\label{sec:results:offset}

Within a single tokamak, the stored-energy response to the same injected neutral-beam power can vary with plasma density, impurity concentration and wall conditioning. For example, lithium wall coatings can alter energy confinement~\cite{maingi2012lithium}, impurity concentrations affect radiative losses~\cite{bernert2017exhaust}, and plasma density influences the fraction of injected beam power that passes through the plasma without being absorbed~\cite{wang2021shine}. A controller that had simply memorised the actuator setpoints for each target in an idealised simulation would reproduce those setpoints for a given target even if the measured stored energy changed. A policy that uses the measured state for feedback should make opposite corrections depending on whether the measured energy is above or below target. To isolate the policy's response to such a shift in the feedback signal, the targets and plant configuration were held fixed and a bias was added only to the stored-energy measurement supplied to the policy.

\begin{figure}[!t]
\centering
\includegraphics{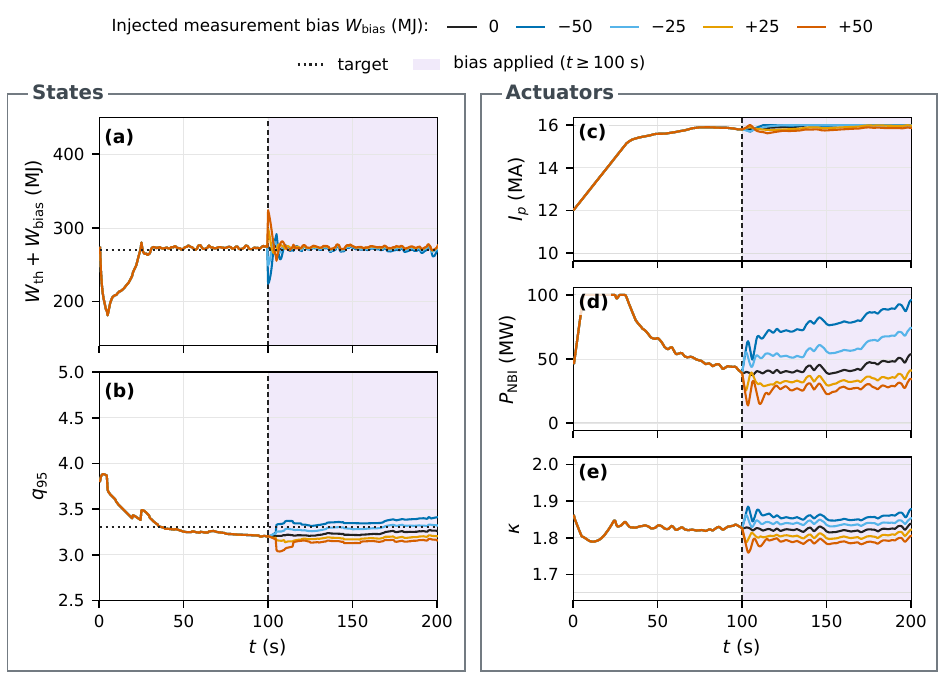}
\caption{Measurement bias test on the training plant at \claim{L306}{$(\Wth^\star,\qff^\star)=(\SI{270}{\mega\joule},3.3)$}. (a) Stored energy supplied to the policy, $\Wth+W_{\mathrm{bias}}$, where $\Wth$ is the energy computed by the plant. At \claim{L306}{\SI{100}{\second}} (vertical dashed line), a constant measurement bias of \claim{L306}{\SI{-50}{\mega\joule}, \SI{-25}{\mega\joule}, \SI{+25}{\mega\joule} or \SI{+50}{\mega\joule}} was applied for the shaded interval. The black line shows the reference case with zero bias throughout. (b) Edge safety factor $\qff$. Dotted lines mark the targets for the two outputs. (c) Plasma current $\Ip$. (d) Neutral-beam power $\Pnbi$. (e) Elongation $\kappa$.}
\label{fig:sensor_offset}
\end{figure}

All \claim{L363}{five} tests used the same random-walk sequences to perturb the plasma parameters and applied actuator values, so the actuator responses to different biases could be compared under a common disturbance history. The target pair was fixed at \claim{L306}{$(\Wth^\star,\qff^\star)=(\SI{270}{\mega\joule},3.3)$}. The reference test (black line in figure~\ref{fig:sensor_offset}) used the unmodified measurement. In each of the other tests, a bias of \claim{L306}{\SI{-50}{\mega\joule}, \SI{-25}{\mega\joule}, \SI{+25}{\mega\joule} or \SI{+50}{\mega\joule}} was added to the stored-energy measurement at \claim{L306}{\SI{100}{\second}} and held constant.

At \claim{L306}{\SI{100}{\second}}, the measured stored energy rose abruptly for positive biases and fell for negative biases. The trajectories of the three actuators then diverged (figure~\ref{fig:sensor_offset}). \claim{L308}{The policy reduced beam power and elongation when the measured energy rose above target and increased both when it fell below target. For biases of the same sign, larger biases produced larger changes in both final actuator values relative to the unbiased case}. The measured stored energy returned close to \claim{L306}{\SI{270}{\mega\joule}}. For all \claim{L363}{five} tests, the terminal errors in measured stored energy and $\qff$ were both within \claim{L363}{5}\,\% of their respective targets (table~\ref{tab:offset}).

\makeatletter
{\setlength{\@fptop}{0pt}\clearpage}
\makeatother

\begin{table}[!htbp]
\caption{Terminal tracking errors in the measurement bias test at \claim{L306}{$(\Wth^\star,\qff^\star)=(\SI{270}{\mega\joule},3.3)$}. $\Delta\Wth^{\rm meas}$ is the signed error in measured stored energy (MJ), and $\Delta\qff$ is the signed safety-factor error. Both are calculated as the target minus the corresponding measurement averaged over the final \claim{L101}{\SI{10}{\second}}. $\epsilon_q$ expresses the safety-factor error as a percentage of its target.}
\centering
\small
\begin{tabular}{c c c c}
\toprule
Bias (MJ) & $\Delta\Wth^{\rm meas}$ (MJ) & $\Delta\qff$ & $\epsilon_q$ (\%) \\
\midrule
none & \claim{L348}{$-2.0$} & \claim{L345}{$+0.039$} & \claim{L345}{$+1.19$} \\
\claim{L307}{$-50$} & \claim{L348}{$+1.9$} & \claim{L345}{$-0.102$} & \claim{L345}{$-3.09$} \\
\claim{L307}{$-25$} & \claim{L348}{$-1.0$} & \claim{L345}{$-0.022$} & \claim{L345}{$-0.67$} \\
\claim{L307}{$+25$} & \claim{L348}{$-2.7$} & \claim{L345}{$+0.097$} & \claim{L345}{$+2.93$} \\
\claim{L307}{$+50$} & \claim{L348}{$-3.3$} & \claim{L345}{$+0.137$} & \claim{L345}{$+4.15$} \\
\bottomrule
\end{tabular}
\label{tab:offset}
\end{table}

The moving target and measurement bias tests show that a single RL policy used both the target and the measured plasma state for feedback. When the stored-energy target changed, \claim{L362}{the policy coordinated plasma current, neutral-beam power and elongation} to reduce the stored-energy error rapidly and then restore $\qff$ tracking. With both targets fixed, the direction and size of the actuator corrections depended on how far the persistently biased energy measurement lay above or below target. The ability to adjust those commands to the measured error could also support tracking when operating conditions change on the same tokamak.

\FloatBarrier

\newpage
\subsection{Plant shift test}
\label{sec:results:transport}

On a newly commissioned tokamak, or one with limited discharge data, the controller may need to bring the plasma to a target operating point before the response to its actuators has been fully characterised. The plant shift test examined how well the same policy could track thermal stored energy $\Wth$ and the edge safety factor $\qff$ without retraining when the plasma response differed from that encountered during training.

\begin{figure}[!t]
\centering
\includegraphics{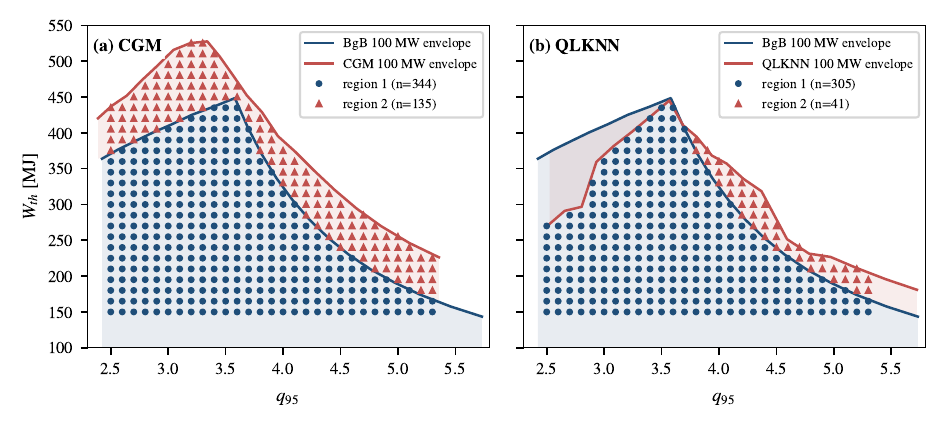}
\caption{Target grids and scan envelopes for the two evaluation plants. (a) CGM. (b) QLKNN. The blue and red curves are the respective \claim{L200}{\SI{100}{\mega\watt}} scan envelopes of the training plant (BgB) and the evaluation plant. Circles mark Region 1 targets within the training domain $\mathcal{T}$ (light blue). Triangles mark Region 2 targets in the area below the evaluation envelope and outside $\mathcal{T}$ (light red).}
\label{fig:region_grid_map}
\end{figure}

The preceding test changed only the stored-energy measurement supplied to the policy. On the other hand, for the plant shift test, the turbulent transport model, pedestal settings, prescribed density profile and absorbed fraction of injected beam power were changed together, altering the plasma response to the actuators. \claim{L367}{At matched settings of plasma current, beam power and elongation, the open-loop scans gave different values of both $\Wth$ and $\qff$ on the training and evaluation plants}. The training plant used BgB. For evaluation, one plant used CGM and the other used QLKNN, a neural-network surrogate for the quasilinear gyrokinetic code QuaLiKiz. Figure~\ref{fig:region_grid_map} shows each plant's scan envelope at a fixed beam power of \claim{L200}{\SI{100}{\mega\watt}} as plasma current and elongation were varied. \claim{L344}{The CGM envelope lay above the BgB envelope in stored energy at every $\qff$ examined, whereas the QLKNN envelope lay below it at low $\qff$}. Targets were selected on or below each evaluation plant's envelope. Region 1 contains targets within the training target domain and Region 2 those outside it.

\FloatBarrier

\begin{figure}[!t]
\centering
\includegraphics{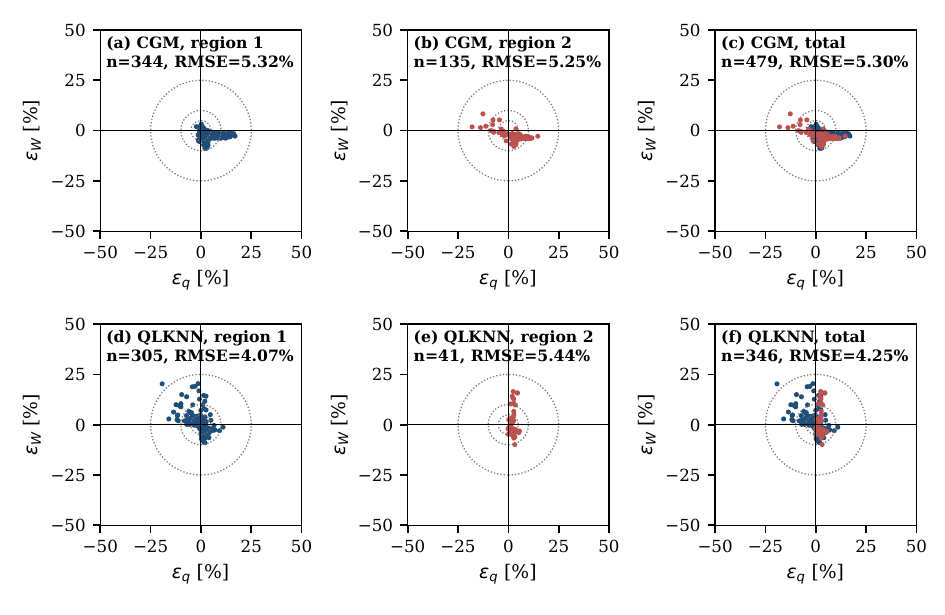}
\caption{Signed relative terminal errors for the target grids in figure~\ref{fig:region_grid_map}. (a) CGM, Region 1. (b) CGM, Region 2. (c) CGM, full grid. (d) QLKNN, Region 1. (e) QLKNN, Region 2. (f) QLKNN, full grid. Dotted circles mark per-target RMSEs of \claim{L347}{5, 10 and 25}\,\%, calculated as $\sqrt{(\epsilon_W^2+\epsilon_q^2)/2}$. Each panel gives the number of targets and the RMSE over both outputs and all targets in that panel.}
\label{fig:region_error_plane}
\end{figure}

Since the plants were evaluated on different target grids (figure~\ref{fig:region_grid_map}), table~\ref{tab:plant_shift_rmse} compares regions within each plant. \claim{L310}{Joint RMSE changed little between regions on CGM and was higher outside the training domain on QLKNN}. At Region~2 targets where beam power stayed at its upper limit throughout the final \claim{L101}{\SI{10}{\second}}, the residual errors differed by plant (figure~\ref{fig:region_error_plane}). \claim{L366}{The CGM cases had high stored-energy targets, with current also at its upper limit and a larger relative RMSE in $\qff$ than in $\Wth$}. \claim{L366}{The QLKNN cases had relatively high safety-factor targets and elongation at its upper limit. Stored energy ended below target while the safety factor was within tolerance}.

\begin{table}[!htbp]
\caption{Joint RMSE (\%) of the relative terminal errors in stored energy and safety factor, calculated as in table~\ref{tab:distribution}. Regions 1 and 2 contain targets inside and outside the training domain, respectively. The full grid includes both regions.}
\centering
\small
\begin{tabular}{l c c c}
\toprule
 & \multicolumn{3}{c}{RMSE (\%)} \\
\cmidrule(lr){2-4}
Evaluation plant & Region 1 & Region 2 & Full grid \\
\midrule
CGM & \claim{L310}{5.32} & \claim{L310}{5.25} & \claim{L310}{5.30} \\
QLKNN & \claim{L310}{4.07} & \claim{L310}{5.44} & \claim{L310}{4.25} \\
\bottomrule
\end{tabular}
\label{tab:plant_shift_rmse}
\end{table}

\FloatBarrier

Even outside the training target domain, \claim{L310}{the joint RMSE remained about 5\,\% on both evaluation plants} without retraining. These results support testing a policy trained in simulation as a starting point for joint control on real tokamak experiments with limited discharge data, before the combination of plasma current, beam power and elongation needed for each target has been mapped experimentally. Tests on such a device could begin at operating points with room to adjust the actuators and then extend towards their limits to identify which output accounts for the residual error.

\FloatBarrier

\section{Conclusion}
\label{sec:conclusion}

Combining integrated tokamak simulation with reinforcement learning yielded a MIMO feedback policy for
tokamak plasmas. In an ITER-scale simulation, one policy
coordinated plasma current, neutral-beam power and elongation to track thermal stored energy $\Wth$
and the edge safety factor $\qff$ across a range of operating points. Four tests examined in detail how the
RL policy used feedback to track the two targets. The policy tracked multiple fixed target pairs,
reached a common target from different
initial states and recovered joint tracking after changes in the stored-energy target. With both
targets held fixed, the direction of its corrections depended on whether the biased stored-energy
measurement was above or below target, and the measurement returned close to target. Under
altered transport and heating conditions, the same policy retained joint tracking without
retraining at some operating points, including targets outside the training domain.

\label{sec:summary:limits}\label{sec:summary:outlook}
The same procedure could guide the development of policies for other plasma quantities and
actuators. Before a new discharge scenario is attempted, a policy can be trained in simulation
and its commands examined at candidate operating points. Using one policy for several targets
could reduce the need to design new actuator waveforms for each initial state and target.
Simulating a scenario under different plasma conditions can reveal how close each actuator comes
to its limit, even at operating points with similar tracking accuracy. For tokamaks with limited
discharge data, this comparison could help identify which operating points to test first and
which plasma responses to examine. Measurements from those discharges could then be used to
check the simulated responses and guide subsequent scenarios over a wider operating range.
Linking policy development in simulation with staged testing in tokamak discharges could provide
a practical route to sim-to-real transfer of RL feedback control.

\begingroup\small
\medskip\noindent{\large\bfseries Acknowledgements}\par\nobreak
\begingroup\raggedright
This research was supported by the Chung-Ang University Graduate Research Scholarship in 2024. This work was supported by a National Research Foundation of Korea (NRF) grant funded by the Korea government (MSIT) (Grant No. RS-2024-00346024). This research was supported by the National Research Foundation (NRF) funded by the Korean government (MSIT) (No. RS-2026-25545529). OpenAI Codex, a generative AI tool, assisted with the writing and English-language editing of this manuscript.\par
\endgroup

\medskip\noindent{\large\bfseries Data availability}\par\nobreak
The data that support the findings of this study are available from the corresponding author upon reasonable request.

\par\endgroup
\clearpage
\begin{appendices}
\hypersetup{bookmarksnumbered=true}
\setcounter{table}{0}\renewcommand{\thetable}{A\arabic{table}}
\renewcommand{\theHtable}{appendix.table.\arabic{table}}
\setcounter{figure}{0}\renewcommand{\thefigure}{A\arabic{figure}}
\renewcommand{\theHfigure}{appendix.figure.\arabic{figure}}
\setcounter{equation}{0}\renewcommand{\theequation}{A.\arabic{equation}}
\renewcommand{\theHequation}{appendix.equation.\arabic{equation}}
\section{Plant parameters, training settings and uncertainty}
\label{sec:appendix:plant}

The parameters below specify the simulation and controller used in sections~\ref{sec:env}
and~\ref{sec:method}. The machine dimensions and magnetic field are common to all configurations,
while composition and density settings vary as shown in table~\ref{tab:machine}. The reference
operating point gives the plasma state about which the actuator sensitivities in
figure~\ref{fig:coupling} were measured; it is not a target imposed throughout the evaluations.
For the boundary in equation~(\ref{eq:boundary}), the geometric triangularity
$(R_0-R_{\rm top})/a$ equals $\sin\delta$. Thus the tabulated Miller coefficient and geometric
triangularity describe the same boundary but are distinct quantities.

\begin{table}[!htbp]
\caption{Machine, composition and density settings of the ITER-scale D--T H-mode simulations. Columns distinguish the training plant (BgB) from the CGM and QLKNN evaluation configurations; merged entries are common to all. The reference block gives the settled training-plant state used for the actuator sensitivity scan. The Miller boundary coefficient $\delta$ corresponds to geometric triangularity $\sin\delta=$ \claim{L349}{0.39}. Density fractions are normalised to the Greenwald density $n_\mathrm{GW}$.}
\centering
\small
\begin{tabular}{l l c c c}
\toprule
Quantity & Symbol & Training (BgB) & CGM & QLKNN \\
\midrule
Major radius            & $R$              & \multicolumn{3}{c}{\claim{L270}{6.2 m}} \\
Minor radius            & $a$              & \multicolumn{3}{c}{\claim{L270}{2.0 m}} \\
Toroidal field          & $B_T$            & \multicolumn{3}{c}{\claim{L270}{5.3 T}} \\
Boundary shaping coefficient & $\delta$ & \multicolumn{3}{c}{\claim{L270}{0.4}} \\
Main ion species        & --               & \multicolumn{3}{c}{\claim{L270}{D:T $=$ 50:50}} \\
Impurity                & --               & \multicolumn{3}{c}{Ne with trace W} \\
\midrule
Effective charge        & $Z_\mathrm{eff}$ & \claim{L270}{2.0} & \claim{L270}{2.0} & \claim{L342}{1.5} \\
Line-averaged density   & $\bar{n}_e$      & \claim{L330}{0.88 $n_\mathrm{GW}$} & \claim{L316}{0.86 $n_\mathrm{GW}$} & \claim{L342}{0.75 $n_\mathrm{GW}$} \\
Separatrix density      & --               & \claim{L270}{0.20 $n_\mathrm{GW}$} & \claim{L270}{0.20 $n_\mathrm{GW}$} & \claim{L342}{0.34 $n_\mathrm{GW}$} \\
\midrule
\multicolumn{5}{l}{Reference operating point (training plant, $\Ip = $ \claim{L264}{15 MA},
$\kappa = $ \claim{L264}{1.88}, $\Pnbi = $ \claim{L264}{50 MW})} \\
Thermal stored energy            & $\Wth$                  & \claim{L264}{306.0 MJ} & -- & -- \\
Edge safety factor               & $\qff$                  & \claim{L264}{3.34} & -- & -- \\
Normalised beta                  & $\beta_N$               & \claim{L330}{1.51} & -- & -- \\
Poloidal beta                    & $\beta_p$               & \claim{L330}{0.46} & -- & -- \\
Energy confinement time          & $\tau_E$                & \claim{L330}{3.02 s} & -- & -- \\
Volume-averaged ion temperature  & $\langle T_i \rangle$   & \claim{L330}{7.7 keV} & -- & -- \\
Central electron temperature     & $T_e(0)$                & \claim{L346}{18.8 keV} & -- & -- \\
Central ion temperature          & $T_i(0)$                & \claim{L346}{17.7 keV} & -- & -- \\
\bottomrule
\end{tabular}
\label{tab:machine}
\end{table}

\FloatBarrier

Table~\ref{tab:numerics} distinguishes the control period from the update intervals for equilibrium geometry and uncertainty. Its setpoint-change bound describes how far the actuator setpoints can
move while the geometry is held. The transport solver uses the stiff-transport stabilisation
scheme of Pereverzev and Corrigan~\cite{pereverzev2008stiff}. The beam deposition shape remains fixed as the policy
changes its power.

\begin{table}[!htbp]
\caption{Timing, discretisation and solver settings shared by the plant configurations. The
equilibrium-update interval determines how long TORAX uses one geometry. The control period and the uncertainty update interval are given separately. Beam deposition parameters
define the fixed heating-profile shape; the absorbed fraction is given separately in
table~\ref{tab:transport}.}
\centering
\begin{tabular}{p{0.42\linewidth} p{0.50\linewidth}}
\toprule
Quantity & Value \\
\midrule
\multicolumn{2}{l}{\emph{Timing}}\\
Transport time step and control period & \claim{L96}{0.5 s} \\
Episode duration & \claim{L96}{200 s (400 control steps)} \\
Equilibrium-update interval & \claim{L202}{every 10 control steps (5 s)} \\
Uncertainty update interval & \claim{L272}{every 4 control steps (2 s)} \\
Largest change of an actuator setpoint within one equilibrium-update interval & \claim{L273}{0.5 MA in $\Ip$; 0.10 in $\kappa$} \\
\multicolumn{2}{l}{\emph{Transport solver}}\\
Solver & \claim{L271}{linear, predictor--corrector (10 corrector steps)} \\
Stiff-term stabilisation & \claim{L271}{Pereverzev--Corrigan ($\chi = 30$ m$^2$\,s$^{-1}$, $D = 15$ m$^2$\,s$^{-1}$)} \\
Radial grid & \claim{L271}{50 cells; 100 flux surfaces from the equilibrium} \\
Outermost surface used as boundary & \claim{L271}{0.99 of the boundary flux} \\
\multicolumn{2}{l}{\emph{Equilibrium solver}}\\
Boundary and solve & prescribed Miller boundary; fixed-boundary solve \\
Poloidal-flux grid & \claim{L271}{\EqGrid} \\
Solve targets & plasma current; on-axis pressure \\
\multicolumn{2}{l}{\emph{Heating source}}\\
Beam deposition location $\hat\rho_\mathrm{dep}$ & \claim{L271}{0.0} \\
Beam deposition width & \claim{L271}{0.3} \\
Electron heating fraction & \claim{L271}{0.7} \\
\bottomrule
\end{tabular}
\label{tab:numerics}
\end{table}

\FloatBarrier

Table~\ref{tab:transport} identifies the heat-transport closure and coefficients for each plant.
The absorbed fraction converts the injected beam power into deposited heating power, so equal
$\Pnbi$ need not give equal heating across configurations. This table describes the transport
and absorption components of the plant change; the prescribed density settings are in
table~\ref{tab:machine}, and the pedestal dependence is described in
section~\ref{sec:env:pedestal}.
The pedestal-width scaling uses the global poloidal beta from the preceding TORAX state in place of
the pedestal-top beta in the original scaling~\cite{onjun2002}.

\begin{table}[!htbp]
\caption{Heat-transport and beam-absorption settings of the training plant (Bohm/gyro-Bohm) and the CGM and QLKNN evaluation configurations. $V_\mathrm{face}$ multiplies the inward convective velocity of the Bohm/gyro-Bohm model; $\alpha_\mathrm{CGM}$ is the exponent of the critical-gradient excess, distinct from the pedestal ballooning gradient $\alpha_c$. ITG, TEM and ETG denote ion-temperature-gradient, trapped-electron and electron-temperature-gradient modes. Unlisted quantities retain the transport-code defaults.}
\centering
\begin{tabular}{l l l}
\toprule
Model & Parameter & Value \\
\midrule
Bohm/gyro-Bohm (training plant) & Bohm coefficient ($\chi_e$, $\chi_i$)       & \claim{L274}{$8\times10^{-5}$} \\
               & gyro-Bohm coefficient ($\chi_e$, $\chi_i$) & \claim{L274}{$5\times10^{-6}$} \\
               & multipliers (all four channels)             & \claim{L274}{2.0} \\
               & convective velocity coefficient $V_\mathrm{face}$ & \claim{L274}{$-0.1$} \\
               & absorbed beam-power fraction & \claim{L275}{0.4} \\
\midrule
Critical gradient (CGM plant) & $\chi_\mathrm{min}$             & \claim{L274}{1.0 m$^2$\,s$^{-1}$} \\
                  & stiffness $\chi_\mathrm{stiff}$ & \claim{L274}{3.0} \\
                  & critical-gradient exponent $\alpha_\mathrm{CGM}$ & \claim{L274}{1.0} \\
                  & $\chi_e/\chi_i$                  & \claim{L274}{1.5} \\
                  & $\chi/D$                         & \claim{L274}{5.0} \\
                  & absorbed beam-power fraction & \claim{L316}{0.85} \\
\midrule
QLKNN (QLKNN plant) & $\chi_\mathrm{min}$ & \claim{L342}{0.75 m$^2$\,s$^{-1}$} \\
                    & ITG, TEM and ETG branches & \claim{L342}{on} \\
                    & absorbed beam-power fraction & \claim{L342}{1.0} \\
\midrule
All & $\chi_\mathrm{max}$ & \claim{L274}{50 m$^2$\,s$^{-1}$} \\
     & smoothing width      & \claim{L274}{0.2} \\
\bottomrule
\end{tabular}
\label{tab:transport}
\end{table}

\FloatBarrier

Table~\ref{tab:training} records the architecture and optimisation settings of the PPO run,
implemented in Stable-Baselines3~\cite{sb3} with PyTorch~\cite{paszke2019pytorch}. The episode budget specifies the requested training duration, and the target-mode
probabilities determine how often the policy encounters stationary and moving requests. Evaluation
reuses the trained policy and frozen observation-standardisation statistics without further
optimisation.

\begin{table}[!htbp]
\caption{Network, sampling and optimisation settings of the reported PPO policy. The episode
budget is the requested training duration, and the target-mode entries are probabilities.
Uncertainty is included throughout training. All evaluations use the policy and
observation-standardisation statistics from this single run without further updates.}
\label{tab:training}
\centering
\small
\setlength{\tabcolsep}{4pt}
\begin{tabular}{l l}
\toprule
Setting & Value \\
\midrule
Algorithm & proximal policy optimisation (PPO)~\cite{ppo} \\
Parallel environments & \claim{L321}{40} \\
Training budget, episodes \(\times\) control steps per episode & \claim{L321}{4000} \(\times\) \claim{L96}{400} \\
Target modes (fixed; step; random walk) & \claim{L92}{0.4; 0.3; 0.3} \\
Discount \(\gamma\) & \claim{L97}{0.99} \\
Policy and value network & multilayer perceptron, hidden layers \claim{L218}{\([256,256]\)} \\
Learning rate & \claim{L218}{\(3\times10^{-4}\)} \\
Rollout per environment, minibatch, epochs & \claim{L215}{256} steps, \claim{L215}{64}, \claim{L215}{10} \\
Clip range & \claim{L215}{0.2} \\
Library versions & \claim{L242}{Stable-Baselines3 2.9.0, PyTorch 2.12.1} \\
\bottomrule
\end{tabular}
\end{table}

\FloatBarrier

\begin{samepage}
Table~\ref{tab:uncertainty} separates the allowed excursion of each random walk from the standard deviation of
an individual Gaussian increment. The bound gives the maximum allowed deviation.
Deviations are expressed as fractions of the nominal parameter values or the actuator commands,
except for the additive deviation in elongation. Uncertainty acts on plant parameters or applied actuator values, while the measurement-bias test independently changes the stored
energy presented to the policy.\par
\end{samepage}

\begin{table}[!htbp]
\caption{Bounded random walks used to model uncertainty in plant parameters and applied actuator values. Bounds and standard deviations of the Gaussian increments are given as fractions of the nominal parameter values or the actuator commands, except for elongation, for which both are absolute. All walks use the update interval in table~\ref{tab:numerics}. Uncertainty in the heat diffusivities applies only to the training plant. The other entries also apply in the CGM and QLKNN evaluations. The symbol $\delta$ denotes the imposed deviation here and the Miller boundary coefficient in equation~(\ref{eq:boundary}).}
\centering
\small
\begin{tabular}{l l r r}
\toprule
Deviation & Quantity & Bound ($\pm$) & Increment standard deviation \\
\midrule
$\delta Z_\mathrm{eff}$ & effective charge & \claim{L278}{0.0432} & \claim{L278}{0.00288} \\
$\delta f_\mathrm{BS}$ & bootstrap-current multiplier & \claim{L278}{0.3195} & \claim{L278}{0.01420} \\
$\delta\chi_e$ & electron heat diffusivity & \claim{L278}{0.0576} & \claim{L278}{0.00288} \\
$\delta\chi_i$ & ion heat diffusivity & \claim{L278}{0.0576} & \claim{L278}{0.00288} \\
$\delta T_\mathrm{ped}$ & applied pedestal temperature & \claim{L278}{0.0360} & \claim{L278}{0.00288} \\
$\delta\Ip$ & applied plasma current & \claim{L278}{0.0344} & \claim{L278}{0.00258} \\
$\delta\Pnbi$ & applied beam power & \claim{L278}{0.0360} & \claim{L278}{0.00288} \\
$\delta\kappa$ & applied elongation (absolute) & \claim{L278}{0.0515} & \claim{L278}{0.00430} \\
\bottomrule
\end{tabular}
\label{tab:uncertainty}
\end{table}

\FloatBarrier

\section{Fixed-target cases outside the joint tolerance}
\label{sec:appendix:fixed_target}
\setcounter{table}{0}\renewcommand{\thetable}{B\arabic{table}}
\renewcommand{\theHtable}{appendix.B.table.\arabic{table}}

For each target outside the joint tolerance in section~\ref{sec:results:nominal},
table~\ref{tab:fixed_target_failures} pairs the signed output errors with the final actuator commands.
The upper bounds on the final commands were \claim{L98}{\SI{16}{\mega\ampere}} for plasma current,
\claim{L98}{\SI{100}{\mega\watt}} for neutral-beam power and \claim{L98}{2.00} for elongation.

\begin{table}[htbp]
\caption{Fixed-target cases outside the joint terminal tolerance on the training plant.
All \claim{L356}{15} cases are listed with their targets, signed relative terminal errors and final actuator setpoints.
Errors follow equation~(\ref{eq:eps}) and use outputs averaged over the final \claim{L101}{\SI{10}{\second}};
a positive error indicates an output below its target. Actuator setpoints are taken from the final control step.
Bold errors exceed \claim{L104}{5}\,\% in magnitude; classification uses unrounded values.}
\centering
\small
\begin{tabular}{r r r r r r r}
\toprule
$\Wth^\star$ (MJ) & $\qff^\star$ & $\epsilon_W$ (\%) & $\epsilon_q$ (\%) & $\Ip$ (MA) & $\Pnbi$ (MW) & $\kappa$ \\
\midrule
\claim{L356}{158.61} & \claim{L356}{5.390} & \textbf{\claim{L356}{+12.71}} & \claim{L356}{-1.66} & \claim{L356}{10.15} & \claim{L356}{100.0} & \claim{L356}{2.000} \\
\claim{L356}{165.25} & \claim{L356}{5.334} & \textbf{\claim{L356}{+12.42}} & \claim{L356}{-1.07} & \claim{L356}{10.22} & \claim{L356}{100.0} & \claim{L356}{2.000} \\
\claim{L356}{168.97} & \claim{L356}{5.274} & \textbf{\claim{L356}{+11.15}} & \claim{L356}{-0.66} & \claim{L356}{10.47} & \claim{L356}{100.0} & \claim{L356}{2.000} \\
\claim{L356}{260.50} & \claim{L356}{2.942} & \claim{L356}{-1.01} & \textbf{\claim{L356}{+6.00}} & \claim{L356}{15.83} & \claim{L356}{21.1} & \claim{L356}{1.768} \\
\claim{L356}{285.80} & \claim{L356}{3.127} & \claim{L356}{-0.94} & \textbf{\claim{L356}{+5.87}} & \claim{L356}{16.00} & \claim{L356}{34.4} & \claim{L356}{1.791} \\
\claim{L356}{301.68} & \claim{L356}{2.776} & \claim{L356}{-0.38} & \textbf{\claim{L356}{+5.01}} & \claim{L356}{16.00} & \claim{L356}{56.9} & \claim{L356}{1.746} \\
\claim{L356}{316.74} & \claim{L356}{2.425} & \claim{L356}{+2.86} & \textbf{\claim{L356}{-5.39}} & \claim{L356}{16.00} & \claim{L356}{100.0} & \claim{L356}{1.665} \\
\claim{L356}{325.03} & \claim{L356}{2.474} & \claim{L356}{+2.90} & \textbf{\claim{L356}{-8.06}} & \claim{L356}{16.00} & \claim{L356}{100.0} & \claim{L356}{1.697} \\
\claim{L356}{363.78} & \claim{L356}{2.603} & \claim{L356}{+2.06} & \textbf{\claim{L356}{-7.78}} & \claim{L356}{16.00} & \claim{L356}{100.0} & \claim{L356}{1.789} \\
\claim{L356}{384.29} & \claim{L356}{3.340} & \textbf{\claim{L356}{+12.16}} & \textbf{\claim{L356}{-8.60}} & \claim{L356}{16.00} & \claim{L356}{100.0} & \claim{L356}{2.000} \\
\claim{L356}{398.06} & \claim{L356}{3.304} & \claim{L356}{+1.96} & \textbf{\claim{L356}{-6.09}} & \claim{L356}{16.00} & \claim{L356}{100.0} & \claim{L356}{2.000} \\
\claim{L356}{401.56} & \claim{L356}{3.098} & \claim{L356}{+2.50} & \textbf{\claim{L356}{-15.21}} & \claim{L356}{16.00} & \claim{L356}{100.0} & \claim{L356}{1.959} \\
\claim{L356}{402.69} & \claim{L356}{3.609} & \textbf{\claim{L356}{+14.28}} & \claim{L356}{-4.27} & \claim{L356}{16.00} & \claim{L356}{100.0} & \claim{L356}{2.000} \\
\claim{L356}{406.20} & \claim{L356}{3.159} & \claim{L356}{+0.97} & \textbf{\claim{L356}{-11.20}} & \claim{L356}{16.00} & \claim{L356}{100.0} & \claim{L356}{1.982} \\
\claim{L356}{414.43} & \claim{L356}{3.411} & \textbf{\claim{L356}{+6.66}} & \textbf{\claim{L356}{-8.81}} & \claim{L356}{16.00} & \claim{L356}{100.0} & \claim{L356}{2.000} \\
\bottomrule
\end{tabular}
\label{tab:fixed_target_failures}
\end{table}

\FloatBarrier

\end{appendices}

\begingroup
\small
\setlength{\bibsep}{2pt plus 0.5pt minus 0.5pt}
\bibliographystyle{iopart-num}
\bibliography{refs}
\endgroup

\end{document}